\documentclass[aip,preprint]{revtex4-1}%linenumbers
\usepackage{graphicx}% Include figure files

\draft % marks overfull lines with a black rule on the right

\begin{document}

%\preprint{}

\title{Numerical calculation of Green's function and momentum distribution for spin-polarized fermions by path integral molecular dynamics}

\author{Yunuo Xiong}
%\email{2111909023@zjut.edu.cn}
\affiliation{College of Science, Zhejiang University of Technology, Hangzhou 31023, China}

\author{Hongwei Xiong}
\email{xionghw@zjut.edu.cn}
\affiliation{College of Science, Zhejiang University of Technology, Hangzhou 31023, China}

\begin{abstract}
Most recently, path integral molecular dynamics (PIMD) has been successfully applied to perform simulations of identical bosons and fermions by B. Hirshberg et al.. In this work, we demonstrate that PIMD can be developed to calculate Green's function and extract momentum distribution for spin-polarized fermions. 
%This work paves the way to study numerically the thermodynamics of fermions by PIMD. 
In particular, we show that the momentum distribution calculated by PIMD has potential application to numerous quantum systems, such as cold atom simulation of Mott insulator in Fermi-Hubbard model.
\end{abstract}


\maketitle

\section{Introduction}

The {\it{ab initio}} simulation of Bose or Fermi systems at finite temperature or zero temperature has a wide range of applications in quantum physics. A particularly important method for this simulation is to write the partition function as a path integral formalism \cite{feynman,kleinert,Tuckerman}, which is further studied by classical ensemble of interacting ring polymers \cite{chandler,Parrinello,Miura,Cao,Cao2,Jang2,Ram,Poly,Craig,Braa,Haber,Thomas}. After this mapping, one may use the Monte-Carlo method \cite{CeperRMP,boninsegni1,boninsegni2,Dornheim} to get huge number of distributions to calculate energy, density distribution, and density-density correlation etc.. The exchange symmetry for bosons and exchange antisymmetry for fermions, however, make the simulation of identical particles a quite challenging work. In addition to path integral Monte-Carlo method, most recently, path integral molecular dynamics (PIMD) for identical particles \cite{Hirshberg,Deuterium,Xiong,HirshbergFermi} is on the rise and is expected to make an increasingly critical contribution to numerous many-particle quantum systems. 

In 2019, PIMD was successfully developed by Hirshberg et al., \cite{Hirshberg} to simulate many-particle Bose system, and this method is applied to study in a remarkable way the supersolid phase in high-pressure Deuterium \cite{Deuterium}. Most recently, we find that the recursion formula in the pioneering work in Ref. \cite{Hirshberg} can be developed to calculate Green's function and momentum distribution \cite{Xiong}, which has been applied successfully to show the Berezinskii-Kosterlitz-Thouless transition \cite{Kosterlitz,Hadzibabic} as an example. Hirshberg and collaborators found in 2020 that PIMD can be also used to study identical fermions without internal degree of freedom \cite{HirshbergFermi}. 

Of course, it would be eager to know whether PIMD can be also used to simulate Green's function for identical fermions, because it is well known that Green's function plays a central role in revealing novel quantum phenomena of fermions \cite{Mahan, Fetter}. For example, the behavior of Green's function for the edge state \cite{Wen} of two-dimensional electrons in a strong magnetic field is the key to understand fractional quantum Hall effect \cite{Tsui} and composite fermion model \cite{Jain}. It is the purpose of this work to develop an efficient method to calculate Green's function for fermions.

In this paper we extend traditional PIMD to study spin-polarized fermions. We show how one can extract Green's function from PIMD simulations, taking the anti-symmetrical wave function for spin-polarized fermions into account. We proceed to consider how to evaluate momentum distribution from Green's function. It is worth noting that momentum distributions are important to experimental studies of quantum many-body systems. Therefore our method has direct applications to cold atom physics. We consider applications of our method to several systems of spin-polarized fermions. In particular, we consider the Mott insulator in Fermi-Hubbard model, which proved crucial to cold atom physics, condensed matter physics and semiconductors.

\section{Partition function and path integral for spin-polarized fermions}

In a recent work \cite{HirshbergFermi} by Hirshberg et al., the recursion formula is given to calculate the partition function of fermions, based on their previous work on bosons \cite{Hirshberg}. Here we give a brief introduction and derivation of the recursion formula, to help us understand how to calculate Green's function for fermions with modified recursion formula.

For the following Hamiltonian operator of $N$ identical spin-polarized fermions
\begin{equation}
\hat H=\frac{1}{2m}\sum_{l=1}^N\hat \textbf p_l^2+\hat V(\textbf r_1,\cdots,\textbf r_N),
\end{equation}
the partition function for spin-polarized fermions is given by
\begin{equation}
Z_F=\frac{1}{N!}\sum_{p\in S_N}(-1)^p\int d\textbf{r}_1d\textbf{r}_2\cdots d\textbf{r}_N\left<p\{\textbf{r}\}|e^{-\beta \hat H}|\{\textbf{r}\}\right>.
\label{Fpartition}
\end{equation}
Here $\beta=1/k_B T$, with $k_B$ being the Boltzmann constant and $T$ being the system temperature. $S_N$ represents the set of $N!$ permutation operations. The factor $(-1)^p$ is due to the exchange effect of fermions. In addition, $\{\textbf{r}\}$ denotes $\{\textbf{r}_1,\cdots,\textbf{r}_N\}$.
%Following the derivation in our previous work \cite{Xiong}, 
Defining normalized position eigenstates for fermions as
\begin{equation}
\left|N_F\right>=\frac{1}{\sqrt{N!}}\sum_{p\in S_N}(-1)^p\left|p\{\textbf{r}\}\right>,
\end{equation}
we have the unit operator
\begin{equation}
\hat I_F=\frac{1}{N!}\int d\textbf{r}_1d\textbf{r}_2\cdots d\textbf{r}_N \left|N_F\right>\left<N_F\right|.
\end{equation}
We also have another unit operator
\begin{equation}
\hat I=\int d\textbf{r}_1d\textbf{r}_2\cdots d\textbf{r}_N \left|\{\textbf{r}\}\right>\left<\{\textbf{r}\}\right|.
\end{equation}

By dividing $\beta$ into $P$ segments, with $\Delta\beta=\beta/P$, and using the identity
\begin{equation}
e^{-\beta \hat H}\hat I_F=e^{-\Delta\beta \hat H}\hat I e^{-\Delta\beta \hat H}\hat I\cdots \hat I e^{-\Delta\beta \hat H}\hat I_F,
\end{equation}
we can obtain the discrete form for the partition function by inserting the definition for the identity operators from the above equations. After that is done, the partition function is a function of a set of ring polymer coordinates $(\textbf{R}_1,\cdots,\textbf{R}_N)$, with $\textbf{R}_i=(\textbf{r}_i^1,\cdots,\textbf{r}_i^P)$ corresponding to $P$ ring polymer coordinates for the $i$th particle. Explicitly, after considering all the permutation terms in Eq. (\ref{Fpartition}), we have \cite{HirshbergFermi}
\begin{equation}
Z_F\sim\int d\textbf{R}_1\cdots d\textbf{R}_N e^{-\beta U_F^{(N)}},
\end{equation}
where $U_F^{(N)}$ is
\begin{equation}
U_F^{(N)}=-\frac{1}{\beta}\ln W_F^{(N)}+\frac{1}{P}\sum_{j=1}^P V\left(\textbf{r}_1^j,\cdots,\textbf{r}_N^j\right),
\label{UFN}
\end{equation}
and $W_F^{(N)}$ is
\begin{equation}
W_F^{(N)}=\frac{1}{N}\sum_{k=1}^N(-1)^ke^{-\beta E_N^{(k)}}W_F^{(N-k)}.
\label{WFN}
\end{equation}
Here the factor $(-1)^k$ accounts for the exchange effect of fermions. For bosons, this factor is always 1. 
\begin{equation}
E_N^{(k)}=\frac{1}{2}m\omega_P^2\sum_{l=N-k+1}^N\sum_{j=1}^P\left(\textbf r_l^{j+1}-\textbf{r}_l^j\right)^2.
\end{equation}
Here $\textbf r_l^{P+1}=\textbf r_{l+1}^1$, except for $l=N$ for which $\textbf r_N^{P+1}=\textbf r_{N-k+1}^1$.
In addition, $\omega_P=\sqrt{P}/\beta\hbar$.

Unlike the corresponding expression for bosons, $W_F^{(N)}$ can be negative and the potential $U_F^{(N)}$ is not real. Therefore we cannot sample the partition function directly through PIMD. 
%Also it is worth noting that the expression for $U_F^{(N)}$ given above leads to a different expression for the fermionic partition function than the one constructed by summing all particle permutations. However, for the purpose of evaluating expectation values they give the same results. 
%For more details about the recursion formula, one may refer to previous works \cite{HirshbergFermi,Hirshberg,Xiong}.
In order to see how we can obtain expectation values for a system of fermions, we denote the estimator for any observable $\hat O$ by $\epsilon_O$ and the expectation value for $\hat O$ is
\begin{equation}
\left<O\right>_F=\frac{1}{Z_F}\int d\textbf{R}_1\cdots d\textbf{R}_N \epsilon_Oe^{-\beta U_F^{(N)}}.
\end{equation}
Here $\left<\cdots\right>_F$ represents the average based on the distribution given by $e^{-\beta U_F^{(N)}}$.We can rewrite $\left<O\right>_F$ as
\begin{equation}
\left<O\right>_F=\frac{1}{Z_F}\int d\textbf{R}_1\cdots d\textbf{R}_N \epsilon_Oe^{-\beta U_B^{(N)}}\frac{e^{-\beta U_F^{(N)}}}{e^{-\beta U_B^{(N)}}},
\end{equation}
with
\begin{equation}
Z_F=\int d\textbf{R}_1\cdots d\textbf{R}_Ne^{-\beta U_B^{(N)}}\frac{e^{-\beta U_F^{(N)}}}{e^{-\beta U_B^{(N)}}},
\end{equation}
where $U_B^{(N)}$ is the potential function for bosons \cite{Hirshberg,Xiong}, which can be obtained from Eqs. (\ref{UFN}) and (\ref{WFN}) without considering the factor $(-1)^k$ in Eq. (\ref{WFN}). $U_B^{(N)}$ is a real function so we can perform PIMD for it. Using the fact that
\begin{equation}
\frac{e^{-\beta U_F^{(N)}}}{e^{-\beta U_B^{(N)}}}=\frac{W_F^{(N)}}{W_B^{(N)}},
\end{equation}
we can obtain the expectation value of any observable quantity for fermionic system as
\begin{equation}
\left<O\right>_F=\frac{\left<\epsilon_O s\right>_B}{\left<s\right>_B},
\end{equation}
where $s=W_F^{(N)}/W_B^{(N)}$. Here $\left<\cdots\right>_B$ represents the average based on the distribution given by $e^{-\beta U_B^{(N)}}$.

For example, the density estimator is given by
\begin{equation}
\rho(\textbf{x})=\left<\frac{1}{P}\sum_{j=1}^P\sum_{k=1}^N\delta(\mathbf r_k^j-\textbf{x})\right>_F.
\end{equation}

As the number of particles increases or the temperature decreases, the magnitude of $\left<s\right>_B$ decreases exponentially, leading to the so called fermion sign problem \cite{ceperley,loh,troyer,lyubartsev,vozn,Yao1,Yao2,Yao3}. It is not the purpose of this work to solve the fermion sign problem. Therefore we consider here only a few fermions so that $\left<s\right>_B$ is not too small to assure the accuracy of our numerical calculations.

\section{PIMD for Green's function and momentum distribution}

The thermal Green's function in this work is defined as \cite{Mahan,Fetter}
\begin{equation}
G(\textbf{x},\tau_1;\textbf{y},\tau_2)=\pm\left<\mathcal T \left\{\hat \psi(\mathbf y,\tau_2)  \hat \psi^\dagger(\mathbf x,\tau_1)\right\}\right>,
\end{equation}
where $\left<\cdots\right>$ denotes thermal average, $\mathcal T$ is the time-ordering operator for field operator. Here, $+$ is for $\tau_2>\tau_1$, while $-$ for $\tau_1>\tau_2$. In addition,
\begin{equation}
\hat \psi(\mathbf x,\tau)=e^{ \hat H \tau}\hat\psi (\mathbf x)e^{-\hat H \tau}, ~~~\hat \psi^{\dagger}(\mathbf x,\tau)=e^{ \hat H \tau}\hat\psi^\dagger  (\mathbf x)e^{-\hat H \tau}.
\end{equation}

From the above Green's function, we may get the momentum distribution of the system. Assuming that $\hat a({\textbf p})$ and 
$\hat a^\dagger({\textbf p})$ are the annihilation and creation operators for a particle with momentum $\textbf p$, we have
\begin{equation}
\hat \psi (\textbf x)=\frac{1}{(2\pi\hbar)^{d/2}}\int d\textbf p \hat a({\textbf p})e^{\frac{i}{\hbar}\textbf p\cdot\textbf x},\hat \psi^\dagger (\textbf x)=\frac{1}{(2\pi\hbar)^{d/2}}\int d\textbf p \hat a^\dagger({\textbf p})e^{-\frac{i}{\hbar}\textbf p\cdot\textbf x}.
\end{equation}
Here $d$ is the spatial dimension of the system. From the above expression, we may also get
\begin{equation}
\hat a({\textbf p})=\frac{1}{(2\pi\hbar)^{d/2}}\int d\textbf x \hat \psi (\textbf x) e^{-\frac{i}{\hbar}\textbf p\cdot\textbf x},
\hat a^\dagger({\textbf p})=\frac{1}{(2\pi\hbar)^{d/2}}\int d\textbf x  \hat \psi^\dagger (\textbf x) e^{\frac{i}{\hbar}\textbf p\cdot\textbf x}.
\end{equation}

The momentum density distribution is
\begin{equation}
\rho(\textbf{p})=\frac{Tr(e^{-\beta\hat H}\hat a^\dagger({\textbf p})\hat a({\textbf p}))}{Tr(e^{-\beta\hat H})}.
\end{equation}
After simple derivations, we get
\begin{equation}
\rho(\textbf{p})=\frac{1}{(2\pi\hbar)^d}\int d{\textbf x}d{\textbf y} G({\textbf x},\tau_1;{\textbf y},\tau_2)
e^{\frac{i}{\hbar}\textbf p\cdot({\textbf x}-{\textbf y})}.
\end{equation}
Here $\tau_1=\tau_2+0^+$.

Hence, in our numerical simulation to get the momentum distribution, we should calculate
\begin{equation}
G(\textbf{x},\tau_2+\Delta\beta;\textbf{y},\tau_2)=\left< \hat \psi^\dagger(\mathbf x,\tau_2+\Delta \beta) \hat \psi(\mathbf y,\tau_2)  \right>.
\end{equation}
Here $\Delta\beta$ is the imaginary time interval of the beads.

\begin{figure}[htbp]
\begin{center}
 \includegraphics[width=0.75\textwidth]{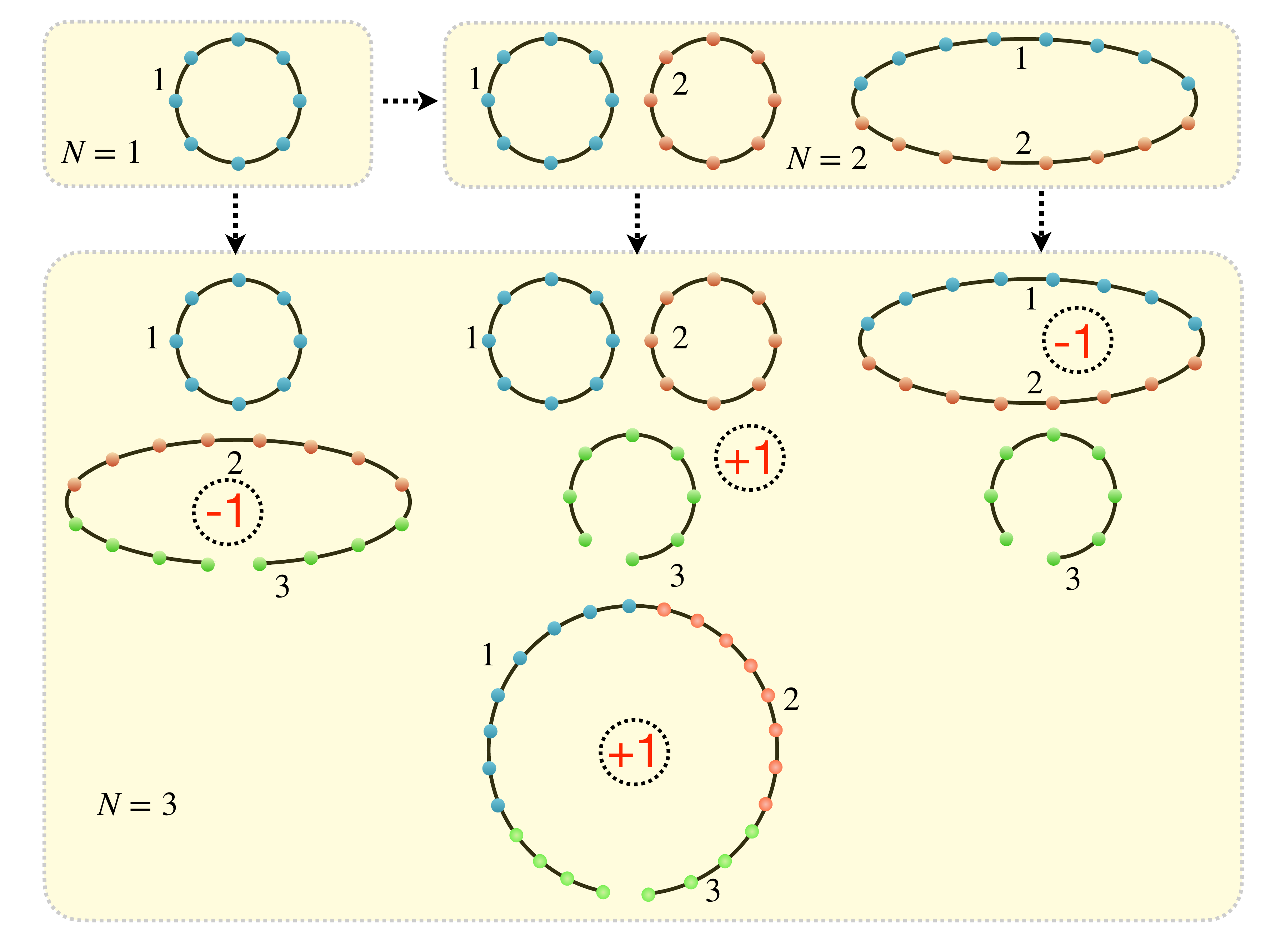} 
\caption{\label{GreenShow} Illustration of all the modified ring polymer configurations for 3 fermions for the case $\tau_1=\tau_2+\Delta \beta$, to calculate Green's function. Different colors have been used to distinguish different ring polymers, green stands for the third ring polymer. The gap indicates that a segment of the third ring polymer is missing. The sign $"+1"$ and $"-1"$ gives the exchange effect of fermions for different situations, respectively.}
\end{center}
\end{figure}

Similarly to Green's function for bosons in our previous work \cite{Xiong}, we may consider the following partition function for Green's function
\begin{equation}
Z_F^G=\int d\mathbf xd\mathbf y Tr\left(e^{-\beta\hat H} \hat\psi^\dagger(\mathbf x,\tau_1) \hat\psi(\mathbf y,\tau_2) \right)
\end{equation}
for $\tau_1=\tau_2+0^+$. 

Repeating the method of the derivation in previous section, we can get the recursion formula for the expression of $Z_F^G$. It is interesting to notice that the method for bosons in Ref. \cite{Xiong} can be generalized to fermions, by including correctly the sign for the permutation of fermions. 
Compared to the boson situation \cite{Xiong}, the main change for the recursion formula to calculate Green's function is the inclusion of the factor $(-1)^k$. 
In Fig. \ref{GreenShow} for $N=3$ fermions, we illustrate the idea of the recursion rule and sign rule for the calculation of Green's function. 

%%%%%%%%%%%%%%%%%
The partition function of Green's function now takes the form
\begin{equation}
Z_F^G\sim \int d\mathbf x d\mathbf y\int d\mathbf R_1\cdots d\mathbf R_{N-1}d\mathbf R_N e^{-\beta (V_G^{(N)}+\frac{1}{P}\sum_{l=1}^P V(\mathbf r_1^l,...,\mathbf r_N^l))}.
\label{PartitionZG}
\end{equation}
Here $\mathbf R_i(1\leq i\leq N-1)$ represents the collection of ring polymer coordinates ($\mathbf r_i^1,\cdots,\mathbf r_i^P$) corresponding to the $i$th ring polymer. $\mathbf R_N$ represents $(\mathbf r_N^1,\cdots, \mathbf r_N^l, \mathbf y,\mathbf x,\mathbf r_N^{l+3},\cdots, \mathbf r_N^{P})$. The gap is between $\mathbf x$ and $\mathbf y$ for the $N$th ring polymer.

\par
In order to modify our PIMD program to accommodate this change, it suffices to modify $E_N^{(k)}$ and its gradient. For the gap considered in this situation, the interaction potential $U_G=\frac{1}{P}\sum_{l=1}^P V(\mathbf r_1^l,...,\mathbf r_N^l))$ between particles will not be changed.
The recursion relation for $V_G^{(\alpha)}$ is
\begin{equation}
e^{-\beta V_G^{(\alpha)}}=\frac{1}{\alpha}\sum_{k=1}^\alpha (-1)^k e^{-\beta(E_\alpha^{(k)}+V_G^{(\alpha-k)})},
\label{recursion2}
\end{equation}
where $V_G^{(0)}=0$. From the above recursion relation, we may get $V_G^{(N)}$ needed in Eq. (\ref{PartitionZG}).
The factor $(-1)^k$ is due to the exchange antisymmetry of fermions.
 %One may refer to the code in GitHub to understand how the algorithm implements $V_G^N$ and $U_G$.

Let us denote positions of the beads corresponding to the $N$th particle at imaginary times $\tau_1(=\tau_2+\Delta \beta)$ and $\tau_2$ as $\mathbf x$ and $\mathbf y$ respectively, we see that for the $N$th particle, $E_N^{(N)}$ should be modified as follows

\begin{eqnarray*}
E_N^{(N)}=&&\frac{1}{2}m\omega_P^2\left\{\left[\sum_{k=1}^{N-1}\sum_{j=1}^P(\mathbf r_k^j-\mathbf r_k^{j+1})^2\right.\right]\\
&&+(\mathbf r_N^1-\mathbf r_N^2)^2+...+(\mathbf r_N^l-\mathbf y)^2\\
&&\left.+(\mathbf x-\mathbf r_N^{l+3})^2+...+(\mathbf r_N^{P}-\mathbf r_N^{P+1})^2\right\},
\end{eqnarray*}
where the boundary conditions remain unaltered. 
%(with $\mathbf r_N^{P+1}$ now equal to $\mathbf r_1^1$). 
The formula for general $E_\alpha^{(k)}$ is the same as before when $\alpha<N$; when $\alpha=N$, it includes all the usual spring energies between beads but without the $(\mathbf x-\mathbf y)^2$ term. 
 
The gradient of $E_N^{(k)}$ should be modified accordingly. For example, the gradient of $E_N^{(N)}$ with respect to $\mathbf x$ is
\begin{equation}
\nabla_{\mathbf x}E_N^{(N)}=m\omega_P^2(\mathbf x-\mathbf r_N^{l+3}).
\end{equation}
%%%%%%%%%%%%%%%%%%%

%With the distribution generated by $Z_F^G$, Green's function for fermions can be estimated as
%\begin{equation}
%G(\textbf{x}',\tau_1;\textbf{y}',\tau_2)=\left<\delta(\textbf{x}-\textbf{x}')\delta(\textbf{y}-\textbf{y}')\right>_F,
%\end{equation}
Of course, similarly to the case of $Z_F$ to calculate the energy and density distribution etc., we should consider the exchange sign in calculating Green's function.
Defining
\begin{equation}
e^{-\beta \tilde{V}_G^{(\alpha)}}=\frac{1}{\alpha}\sum_{k=1}^\alpha  e^{-\beta(E_\alpha^{(k)}+\tilde{V}_G^{(\alpha-k)})}
\label{recursion3}
\end{equation}
and
\begin{equation}
\tilde{Z}_G \sim \int d\mathbf x d\mathbf y\int d\mathbf R_1\cdots d\mathbf R_{N-1}d\mathbf R_N e^{-\beta (\tilde{V}_G^{(N)}+\frac{1}{P}\sum_{l=1}^P V(\mathbf r_1^l,...,\mathbf r_N^l))},
\label{PartitiontZG}
\end{equation}
Green's function is
\begin{equation}
G(\textbf{x}',\tau_2+\Delta\beta;\textbf{y}',\tau_2)=\frac{\left<\delta(\textbf{x}-\textbf{x}')\delta(\textbf{y}-\textbf{y}')
e^{\beta(\tilde{V}_G^{(N)}-{V}_G^{(N)})}\right>_{{\tilde Z}_G}}{\left<e^{\beta(\tilde{V}_G^{(N)}-{V}_G^{(N)})}\right>_{{\tilde Z}_G}}.
\end{equation}
Here $\textbf{x}$ and $\textbf{y}$ denote the positions of two beads at the end of the gap.  $\left<\cdots\right>_{{\tilde Z}_G}$ denotes the average for the distribution generated by Eq. (\ref{PartitiontZG}).  The denominator in the above equation can be used to monitor the validity of our simulation.

\section{Examples to implement our algorithm}

In order to test our method and show how to apply our method to realistic quantum system, we apply our algorithm to study various systems of identical fermions. In all of the following simulations,  we will use massive Nos\'e-Hoover chain \cite{Nose1,Nose2,Hoover,Martyna,Jang} to establish constant temperature for the system, where each degree of freedom of the system has been coupled to a separate Nos\'e-Hoover thermostat. 
The number of beads used decreases as temperature increases to ensure numerical stability and assure convergence.
In all of the following we checked convergence with respect to the number of beads and MD steps performed. For details of how to assure the convergence, one may refer to the supplementary material in Ref. \cite{Hirshberg} All the parameters chosen in the following calculations will not lead to severe fermion sign problem that influences the accuracy of our simulations.

\subsection{Ideal fermions in a harmonic trap}

We first consider two noninteracting fermions in a two-dimensional harmonic trap. The Hamiltonian operator of the system is
\begin{equation}
\hat H=-\frac{1}{2}\left(\Delta_1+\Delta_2\right)+\frac{1}{2}({\bf{r}} _1^2+{\bf{r}}_2^2).
\end{equation}
Here we have used the units with $\hbar=1,m=1,\omega=1$.

To test our numerical simulations, we first give the result by other method for comparison. In grand canonical ensemble, the average occupation number $N_{n_xn_y}$ is
\begin{equation}
N_{n_xn_y}=\frac{1}{e^{\beta(\epsilon_{n_xn_y}-\mu)}+1},~~~n_x,n_y=0,1,\cdots
\end{equation}
Here the single-particle energy spectrum $\epsilon_{n_xn_y}=n_x+n_y+1$.

For a given $\beta$ (in unit of $(\hbar\omega)^{-1}$), we may use the following equation to determine the chemical potential $\mu$.
\begin{equation}
N=\sum_{n_xn_y}\frac{1}{e^{\beta(\epsilon_{n_xn_y}-\mu)}+1},
\end{equation}
with $N$ the total number of particles.

The average density distribution in grand canonical ensemble is then
\begin{equation}
\rho(\textbf{r})=\sum_{n_xn_y}\frac{1}{e^{\beta(\epsilon_{n_xn_y}-\mu)}+1}|\psi_{n_x n_y}(\textbf{r})|^2.
\end{equation}
Here $\psi_{n_x n_y}(\textbf{r})=\psi_{n_x}(x)\psi_{n_y}(y)$ 
is the eigen wavefunction for the energy eigenvalue $\epsilon_{n_xn_y}$, and
\begin{equation}
\psi_n(x)=\left(\frac{1}{\pi}\right)^{1/4}\frac{1}{\sqrt{2^n n!}}H_n(x)e^{-\frac{x^2}{2}},
\end{equation}
with $H_n(x)$ the Hermite polynomials.

The average density distribution along $x$ direction is
\begin{equation}
\rho(x)=\int_{-\infty}^{\infty}dy \rho(\textbf{r})=\sum_{n_xn_y}\frac{1}{e^{\beta(\epsilon_{n_xn_y}-\mu)}+1}|\psi_{n_y}(x)|^2.
\end{equation}

\begin{figure}[htbp]
\begin{center}
 \includegraphics[width=0.75\textwidth]{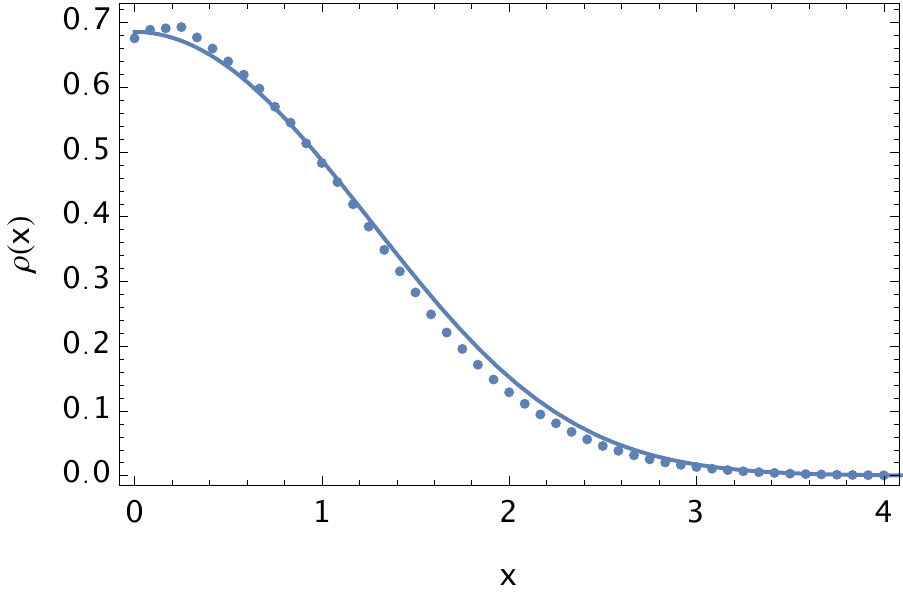} 
\caption{\label{IdealDensity} Density for 2 fermions in a two-dimensional harmonic trap with $\beta=1$. The solid line is the result of grand canonical ensemble, while the solid circle is the simulation result of PIMD ($12$ beads and $2\times10^7$ MD steps to assure convergence). }
\end{center}
\end{figure}

In Fig. \ref{IdealDensity}, the solid line shows $\rho(x)$ for $x>0$ with the above equation for $\beta=1$. The solid circle shows our numerical result of the density distribution with PIMD, which shows good agreement.

\begin{figure}[htbp]
\begin{center}
 \includegraphics[width=0.75\textwidth]{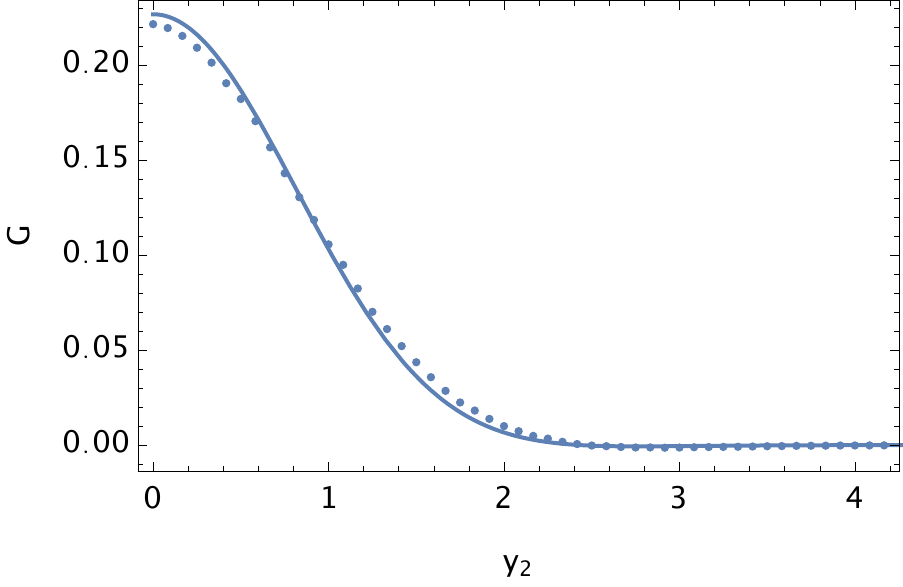} 
\caption{\label{IdealGreen} Green's function for 2 fermions in a harmonic trap with $\beta=1$. The solid line is the result of grand canonical ensemble, while the solid circle is the simulation result of PIMD ($12$ beads and $2\times10^7$ MD steps to assure convergence).}
\end{center}
\end{figure}

In grand canonical ensemble, Green's function for $\tau_1=\tau_2+0^+$ is 
\begin{equation}
G(\textbf{r}_1,\textbf{r}_2)=\sum_{n_xn_y}\frac{1}{e^{\beta(\epsilon_{n_xn_y}-\mu)}+1}
\psi_{n_x n_y}(x_1,y_1)\psi_{n_x n_y}(x_2,y_2).
\end{equation}
In Fig. \ref{IdealGreen}, the solid line shows Green's function $G(x_1=0,y_1=0,x_2=0,y_2)$ for $y_2>0$ in grand canonical ensemble. In the same figure, the solid circle gives the numerical result based on PIMD, which shows good agreement. The slight difference is due to the fact that what the PIMD simulates is canonical ensemble, while the solid lines in Fig. \ref{IdealDensity} and Fig. \ref{IdealGreen} are the result of grand canonical ensemble.

\subsection{Green's function and momentum distribution for interacting fermions}

Now we turn to consider spin-polarized electrons with Coulomb interaction without external trap. The Hamiltonian operator is
\begin{equation}
\hat H=-\frac{1}{2}\sum_{j=1}^N\Delta_j+\sum_{k>l}^N \frac{1}{\left|\textbf{r}_k-\textbf{r}_l\right|}.
\label{Htrap}
\end{equation}
%This Hamiltonian operator has application to quantum dot, etc..

\begin{figure}[htbp]
\begin{center}
 \includegraphics[width=0.75\textwidth]{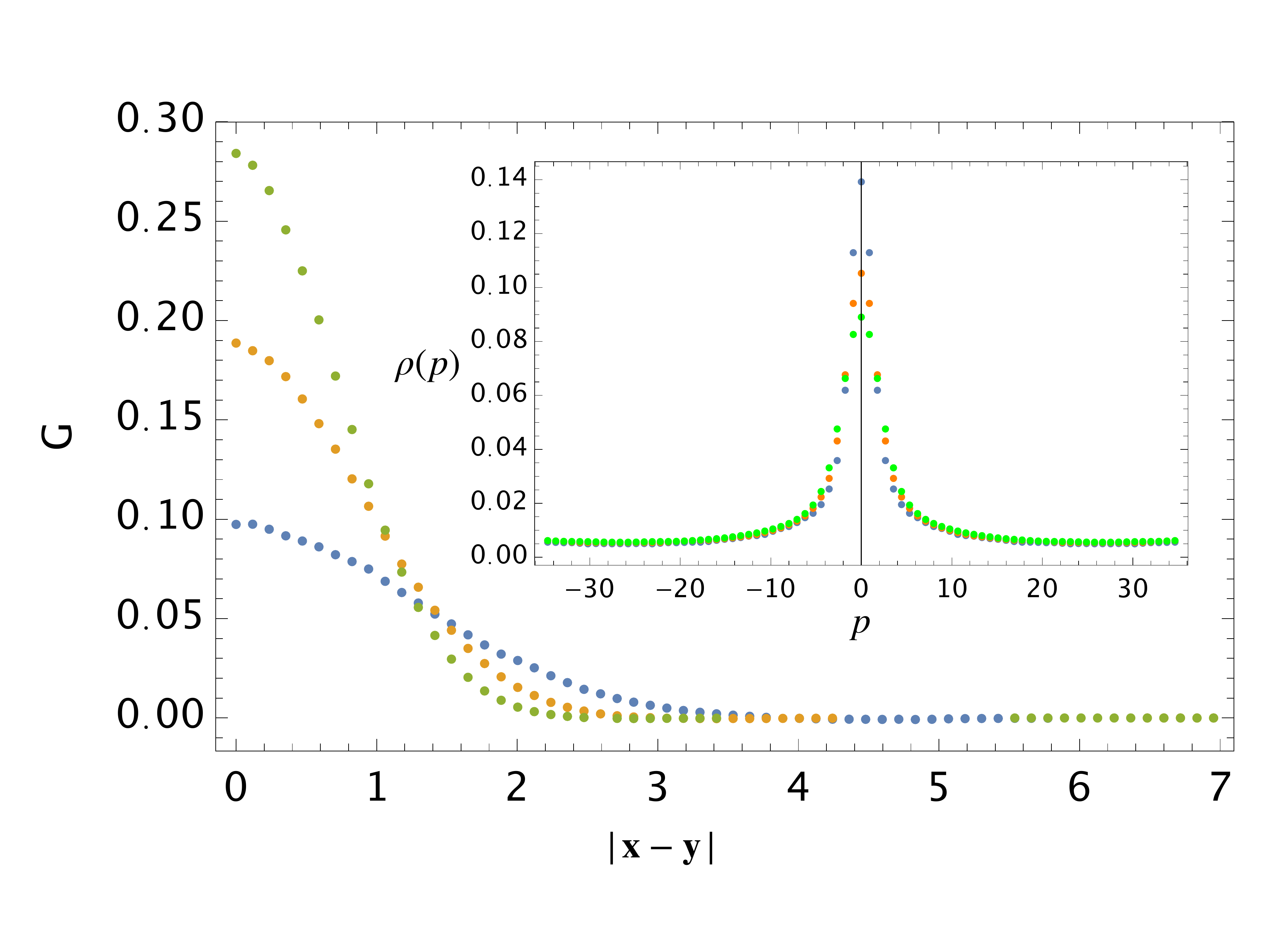} 
\caption{\label{interactingE} Green's function for 4 fermions without external trap. Green corresponds to Green's function for $\beta=0.5$ (8 beads), while orange for $\beta=1.0$ (12 beads), blue for $\beta=1.5$ (24 beads). The inset shows the momentum distribution for corresponding temperatures.}
\end{center}
\end{figure}

We adopt periodic boundary condition with $L=10$ and $4$ particles in our simulation.
In Fig. \ref{interactingE}, we give the numerical result of Green's function by PIMD for different temperatures. $2\times10^7$ MD steps are used to assure convergence. 
%In addition, the number of beads used decreases proportionally as temperature increases to ensure numerical stability and assure convergence, so that $\Delta\beta$ is the same for different temperatures. 
We see that with increasing temperature, Green's function becomes narrower, and we expect that the momentum width would be wider for higher temperature. 

For a uniform system, we have
\begin{equation}
G(\textbf{x},\tau_1;\textbf{y},\tau_2)=G(\textbf{x}-\textbf{y},\tau_1-\tau_2).
\end{equation}
In this case, the momentum distribution becomes
\begin{equation}
\rho(\textbf p)=\frac{L^d}{(2\pi\hbar)^d}\int d{\textbf x} G({\textbf x},\tau_1-\tau_2)
e^{\frac{i}{\hbar}\textbf p\cdot{\textbf x}},~\tau_1=\tau_2+0^+.
\end{equation}
In the inset of Fig. \ref{interactingE}, we show the momentum distribution for different temperatures, which verifies our expectation. For spin-polarized electrons, we see that there is a continuous change with decreasing temperature.

\subsection{Momentum distribution for ultracold fermionic atoms in a harmonic trap}

We consider here an example of the following Hamiltonian operator which may have potential applications in  ultracold fermionic atoms.

\begin{equation}
\hat H=-\frac{1}{2}\sum_{j=1}^N\Delta_j+\sum_{k>l}^N \frac{\lambda}{\left|\textbf{r}_k-\textbf{r}_l\right|^3}
+\frac{1}{2}\sum_{k=1}^N\textbf{r}_k^2.
\label{Htrap}
\end{equation}
In the above $\hat H$, we consider dipole-dipole interaction \cite{Stuhler,Filinov} between atoms, which is a typical and interesting interaction in cold atom physics.
%We may consider different $\lambda$ to consider this problem.

%\begin{equation}
%\hat H=-\frac{1}{2}\sum_{j=1}^N\Delta_j+\sum_{j=1}^N\frac{1}{2}\textbf{r}_j^2+\sum_{k>l}^N\frac{g}{\pi s^2}e^{-\frac{(\mathbf r_k-\mathbf r_l)^2}{s^2}}.
%\label{Hmott}
%\end{equation}

Because of the presence of the harmonic trap, Green's function can not be written as a function of $\mathbf r_1-\mathbf r_2$. In the unit $\hbar=1$, the momentum distribution in two dimensions is
\begin{equation}
\rho(p_x,p_y)=\frac{1}{(2\pi)^2}\int d{\mathbf r_1}d{\mathbf r_2} G({\mathbf r_1},\tau_2+0^+;{\mathbf r_2},\tau_2)
e^{i [p_x(x_1-x_2)]+p_y(y_1-y_2)]}.
\end{equation}
The momentum distribution in $x$ direction is then
\begin{equation}
\rho(p_x)=\int dp_y \rho(p_x,p_y).
\end{equation}
We have
\begin{equation}
\rho(p_x)=\frac{1}{2\pi}\int dy_1 dx_1dx_2G(x_1,y_1,\tau_2+0^+;x_2,y_1,\tau_2)e^{ip_x(x_1-x_2)}.
\end{equation}
It is easy to verify that $\int \rho(p_x)dp_x=N$. From the above equation, we may calculate $\rho(p_x)$ from the numerical simulation of Green's function $G(x_1,y_1,\tau_2+0^+;x_2,y_1,\tau_2)$ by PIMD. 

\begin{figure}[htbp]
\begin{center}
 \includegraphics[width=0.75\textwidth]{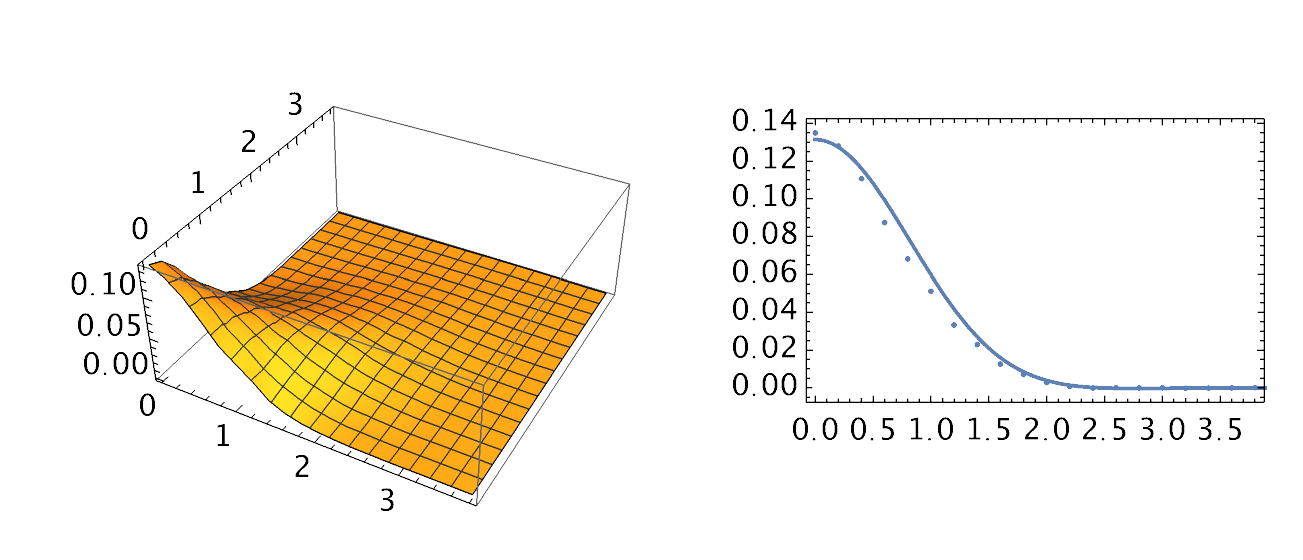} 
\caption{\label{Green3d} 3D Green's function as a function of $x_1>0$ and $x_2>0$ for 2 fermions in harmonic trap without interaction. In the right figure we take a cross section of Green's function and compare it with the result of grand canonical ensemble.}
\end{center}
\end{figure}

In this case, to get the momentum distribution for nonuniform system, we had to calculate Green's function with multiple variables. In Fig. \ref{Green3d}, we show the following Green's function for 2 fermions in a harmonic trap without interaction. We find good agreement with the result of grand canonical ensemble.
\begin{equation}
G(x_1,\tau_2+\Delta\beta; x_2,\tau_2)=\int dy_1 G(x_1,y_1,\tau_2+\Delta\beta;x_2,y_1,\tau_2).
\end{equation}

\begin{figure}[htbp]
\begin{center}
 \includegraphics[width=0.75\textwidth]{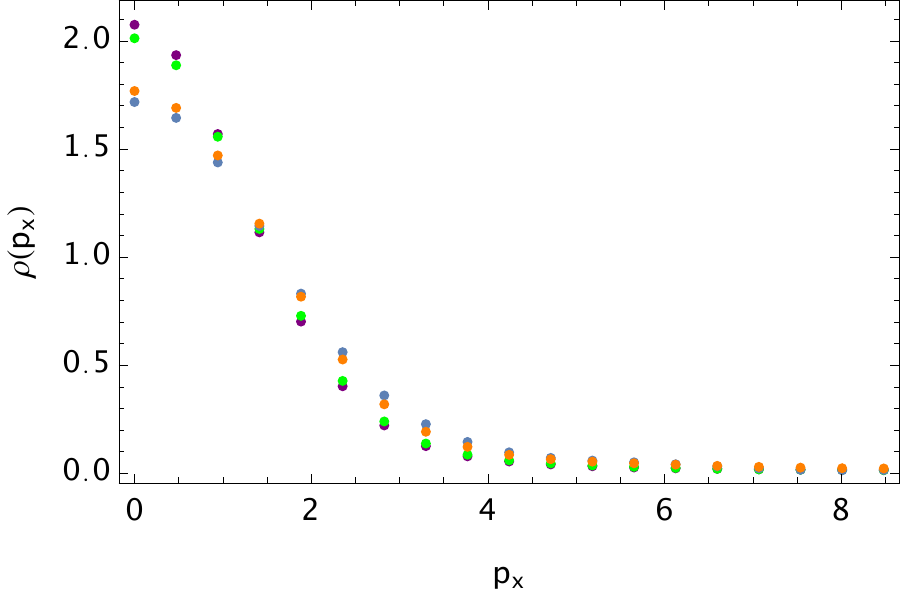} 
\caption{\label{Greenmomentum} Momentum distribution for $N=4$, $\beta=1$ and $\lambda=0.1$ (blue), 0.5 (orange),10 (green),15 (purple).}
\end{center}
\end{figure}

In Fig. \ref{Greenmomentum},  we show the momentum distribution for $N=4$, $\beta=1$ and different $\lambda$. Because of the fermion sign problem, we can not yet simulate accurately the case of a large number of fermions. However, even for a few fermions, our simulation may have direct application. In cold atom experiments, usually thousands of ultracold fermionic atoms are studied. However, in many remarkable experiments, periodic optical lattice is used to simulate the Fermi-Hubbard model \cite{Esslinger}. In this case, in the Mott insulator state, for each lattice site, there are only a few fermions confined in an equivalently harmonic trap.  After switching off the optical lattice and the confining trap, what the experiment measures directly is the momentum distribution of the fermionic atoms in each lattice site, after sufficient time of free expansion \cite{Dalfovo,Anderson,Davis}. The usual Fermi-Hubbard model can not calculate accurately the momentum distribution of the fermionic atoms in each lattice site because the wave function is usually approximated by noninteracting gas, while the method developed here can provide accurate calculation of the momentum distribution. Hence, our method may have direct application in the cold atom simulation of the Fermi-Hubbard model. Of course, this means that the Bose version of our method will have potential application for cold atom simulation of Bose-Hubbard model \cite{Greiner}.

%After switching off all the trapping potential and after a sufficiently long time of free expansion, what the experiment measures is the momentum distribution in the presence of the optical lattice. 

\subsection{Momentum distribution for different statistics}

\begin{figure}[htbp]
\begin{center}
 \includegraphics[width=0.75\textwidth]{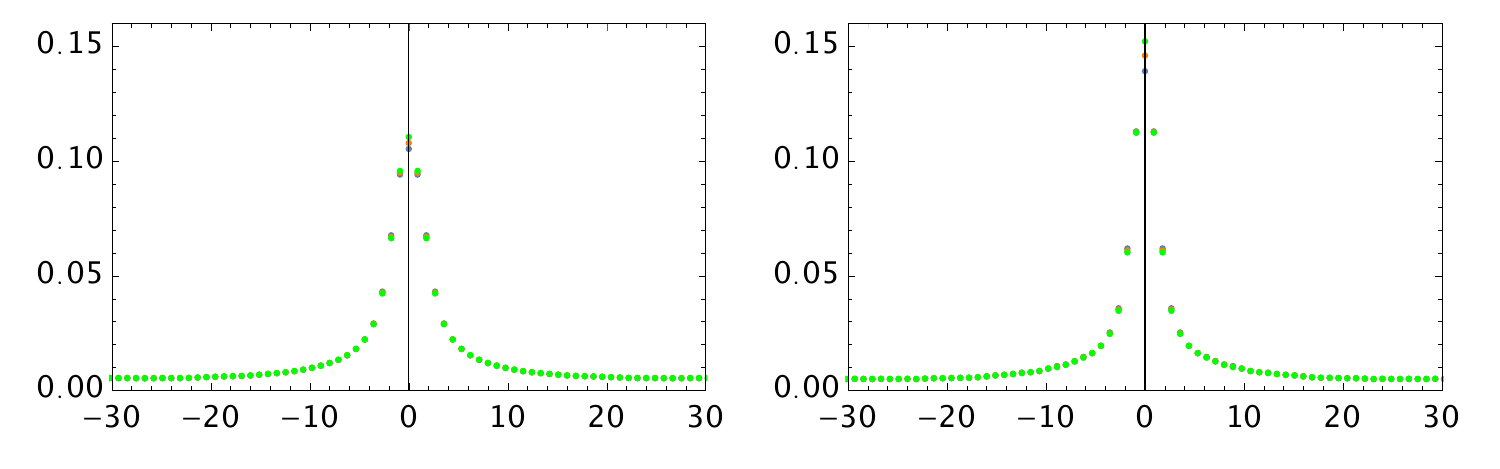} 
\caption{\label{Dstatistics} Momentum distribution for 4 fermions (blue)/distinguishable particles (orange)/bosons (green) with $1/r$ interaction. The left figure is for $\beta=1.0$, while the right figure is for $\beta=2.0$.}
\end{center}
\end{figure}

Finally, as a comparison, we consider the momentum distribution 
of 4 particles with the same $\hat H$ given by Eq. (\ref{Htrap}) for fermions, bosons and distinguishable particles, respectively. In Fig. \ref{Dstatistics}, we show the momentum distribution for different statistics for $\beta=1$ and 
$\lambda=1$. We see that the bosons have the largest peak value, while the fermions have the smallest peak value, which is expected from the general consideration of statistical interaction. However, for this temperature, the difference between different statistics is not obvious. It is expected that decreasing the temperature would make the difference between different statistics more obvious. In the right figure, we give the momentum distribution for $\beta=2.0$. We do see that the difference becomes more obvious. We also calculate the situation without interactions, and find that the difference of the momentum distributions for bosons and fermions becomes larger, for the same temperature. 

\section{Conclusions}

As a summary, we extended PIMD method to calculate Green function and momentum distribution for spin-polarized fermions. We applied our method to study several systems of spin-polarized fermions. In particular, we studied the realistic momentum distribution for a few fermions in Mott insulator, which are of interests to experimental cold atoms physics for Fermi-Hubbard model. Our method can easily be adapted to study numerous realistic quantum systems.

\begin{acknowledgments}
This work
is partly supported by the National Natural Science Foundation of China under grant numbers 11175246, and 11334001.
\end{acknowledgments}

\textbf{DATA AVAILABILITY}

The data that support the findings of this study are available from the corresponding author upon reasonable request.
The code of this study is openly available in GitHub (https://github.com/xiongyunuo/PIMD-for-Green-Function-for-Fermions).


\begin{thebibliography}{10}

%path integral book
\bibitem{feynman} R. P.~Feynman and A. R.~Hibbs, Quantum mechanics and path integrals, Dover Publications, New York (2010).

\bibitem{kleinert} H.~Kleinert, Path integrals in quantum mechanics, statistics, polymer physics, and financial markets, World Scientific, Singapore (2009).

\bibitem{Tuckerman} M. E.~Tuckerman, Statistical mechanics: theory and molecular simulation, Oxford University, New York (2010).


%Ring polymer
\bibitem{chandler} D.~Chandler and P. G.~Wolynes, Exploiting the isomorphism between quantum theory and classical statistical mechanics of polyatomic fluids, {\text{J.~Chem.~Phys.}~\textbf{74}, 4078} (1981).

\bibitem{Parrinello} M.~Parrinello and A.~Rahman, Study of an F center in molten KCl, \text{J. Chem. Phys.}~\textbf{80}, 860 (1984).

\bibitem{Miura} S. Miura and S. Okazaki, Path integral molecular dynamics for Bose-Einstein and Fermi-Dirac statistics. \text{J. Chem. Phys.} \textbf{112}, 10116 (2000).

 \bibitem{Cao} J. Cao and G. A. Voth,  The formulation of quantum statistical mechanics based on the Feynman path centroid density. I. Equilibrium properties, \text{J. Chem. Phys.} \textbf{100},  5093 (1994).
 
 \bibitem{Cao2} J. Cao and G. A. Voth, The formulation of quantum statistical mechanics based on the Feynman path centroid density. II. Dynamical properties,  \text{J. Chem. Phys.} \textbf{100}, 5106 (1994).
 
 \bibitem{Jang2} S. Jang and G. A. Voth, A derivation of centroid molecular dynamics and other approximate time evolution methods for path integral centroid variables,  \text{J. Chem. Phys.} \textbf{111}, 2371 (1999).
 
\bibitem{Ram} R. Ram\'iRez and T. L\'oPez-Ciudad, The Schr\"odinger formulation of the Feynman path centroid density, \text{J. Chem. Phys.} \textbf{111}, 3339 (1999).
 
\bibitem{Poly} E. A. Polyakov, A. P.  Lyubartsev, and P. N.  Vorontsov-Velyaminov,  Centroid molecular dynamics: Comparison with exact results for model systems, \text{J. Chem. Phys.} \textbf{133}, 194103 (2010). 

 \bibitem{Craig} I. R. Craig and D. E. Manolopoulos, Quantum statistics and classical mechanics: Real time correlation functions from ring polymer molecular dynamics,  \text{J. Chem. Phys.}  \textbf{121}, 3368 (2004). 
 
 \bibitem{Braa} B. J. Braams and D. E. Manolopoulos, On the short-time limit of ring polymer molecular dynamics, \text{J. Chem. Phys.} \textbf{125}, 124105 (2006).
 
 \bibitem{Haber} S. Habershon, D. E. Manolopoulos, T. E. Markland, and T. F. Miller 3rd, Ring-polymer molecular dynamics: quantum effects in chemical dynamics from classical trajectories in an extended phase space, \text{Annu. Rev. Phys. Chem.} \textbf{64}, 387 (2013).
 
 \bibitem{Thomas} T. E. Markland and M. Ceriotti, Nuclear quantum effects enter the mainstream, \text{Nat. Rev. Chem.} \textbf{2}, 0109 (2018).
 
 %Path integral Monte Carlo
\bibitem{CeperRMP} D. M. Ceperley, Path integrals in the theory of condensed helium, \text{Rev. Mod. Phys.} \textbf{67}, 279 (1995).

\bibitem{boninsegni1} M.~Boninsegni, N. V.~Prokof’ev, and B. V.~Svistunov, Worm algorithm and diagrammatic Monte Carlo: A new approach to continuous-space path integral Monte Carlo simulations,  {\text{Phys.~Rev.~E}~\textbf{74}, 036701} (2006).

\bibitem{boninsegni2} M.~Boninsegni, N. V.~Prokof’ev, and B. V.~Svistunov, Worm algorithm for continuous-space path integral Monte Carlo simulations,  {\text{Phys.~Rev.~Lett.}~\textbf{96}, 070601} (2006).

\bibitem{Dornheim} T.~Dornheim, The Fermion sign problem in path integral Monte Carlo simulations: quantum dots, ultracold atoms, and warm dense matter, \text{Phys. Rev. E}~\textbf{100}, 023307 (2019).

%Parrinello bosons
\bibitem{Hirshberg} B. Hirshberg, V. Rizzi, and M. Parrinello, Path integral molecular dynamics for bosons, \text{Proc. Natl. Acad. Sci. U. S. A.}~\textbf{116}, 21445 (2019).

%supersolid phase in high-pressure deuterium bosonic system
\bibitem{Deuterium}   C. W. Myung, B. Hirshberg, and M. Parrinello, 
Prediction of a supersolid phase in high-pressure deuterium, \text{Phys. Rev. Lett.} \textbf{128}, 045301 (2022).

%our work about Green's function

\bibitem{Xiong} Y. N. Xiong and H. W. Xiong, Path integral molecular dynamics simulations for Green's function in a system of identical bosons, arXiv:2203.09919 (2022). (accepted by The Journal of Chemical Physics)

%Parrinello fermions
\bibitem{HirshbergFermi} B. Hirshberg,  M. Invernizzi, and  M. Parrinello, Path integral molecular dynamics for fermions: Alleviating the sign problem with the Bogoliubov inequality, \text{J. Chem. Phys.} \textbf{152}, 171102 (2020).

%KT transition
\bibitem{Kosterlitz} J. M. Kosterlitz and D. J. Thouless, Metastability and phase transitions in two dimensional systems, \text{J. Phys. C} \textbf{6}, 1181 (1973).

\bibitem{Hadzibabic} Z. Hadzibabic, P. Kr\"uger, M. Cheneau, B. Battelier, and J. Dalibard, Berezinskii-Kosterlitz-Thouless crossover in a trapped atomic gas, \text{Nature} \textbf{41}, 1118 (2006).


%GreeFunction
\bibitem{Mahan} G. D. Mahan, Many-particle physics, Plenum, New York (2000).

\bibitem{Fetter} A. L. Fetter and J. D. Walecka, Quantum theory of many-particle systems, McGraw-Hill, New York (1971).

%Edge state

\bibitem{Wen} X. G. Wen,Theory of the edge states in fractional quantum Hall effects, Int. J. Mod. Phys. B \textbf{6}, 1711 (1992).

% fractional quantum Hall effect

\bibitem{Tsui} D. C. Tsui, H. L. Stormer, and A. C. Gossard, Two-Dimensional Magnetotransport in the Extreme Quantum Limit,
\text{Phys. Rev. Lett.} \textbf{48}, 1559 (1982).

% composite fermion model

\bibitem{Jain} K. J. Jainendra,  Composite fermions, Cambridge University Press, Cambridge, (2007).

%Fermion sign problem

\bibitem{ceperley} D. M.~Ceperley, Path Integral Monte Carlo Methods for Fermions, Monte Carlo and Molecular Dynamics of Condensed Matter Systems, K.~Binder and G.~Ciccotti (Eds.), Bologna (Italy) (1996).

\bibitem{loh} E. Y.~Loh, J. E.~Gubernatis, R. T.~Scalettar, S. R.~White, D. J.~Scalapino, and R. L.~Sugar, Sign problem in the numerical simulation of many-electron systems, {\text{Phys. Rev. B} \textbf{41}, 9301} (1990).

\bibitem{troyer} M.~Troyer and U. J.~Wiese, Computational Complexity and Fundamental Limitations to Fermionic Quantum Monte Carlo Simulations, {\text{Phys. Rev. Lett.} \textbf{94}, 170201} (2005).

\bibitem{lyubartsev} A. P.~Lyubartsev, 
Simulation of excited states and the sign problem in the path integral Monte Carlo method, {\text{J.~Phys.~A: Math.~Gen.}~\textbf{38}, 6659} (2005).

\bibitem{vozn} M. A.~Voznesenskiy, P. N.~Vorontsov-Velyaminov, and A. P.~Lyubartsev, Path-integral-expanded-ensemble Monte Carlo method in treatment of the sign problem for fermions, \text{Phys.~Rev.~E}~\textbf{80}, 066702 (2009).

\bibitem{Yao1} Z. X.~Li, Y.~F. Jiang, and H.~Yao, Solving the fermion sign problem in quantum Monte Carlo simulations by Majorana representation, {\text{Phys.~Rev.~B} \textbf{91}, 241117(R)} (2015).

\bibitem{Yao2} Z. X.~Li, Y.~F. Jiang, and H.~Yao, Majorana-Time-Reversal Symmetries: A Fundamental Principle for Sign-Problem-Free Quantum Monte Carlo Simulations, { \text{Phys.~Rev.~Lett.}~\textbf{117}, 267002} (2016).

\bibitem{Yao3} Z. X.~Li and H.~Yao, Sign-Problem-Free Fermionic Quantum Monte Carlo: Developments and Applications, { \text{Ann.~Rev.~Cond.~Mat.~Phys.}~\textbf{10}, 337} (2019).


%Nose-Hoover chains

\bibitem{Nose1} S. Nos\'e, A molecular dynamics method for simulations in the canonical ensemble, \text{Mol. Phys.} \textbf{52}, 255 (1984).

\bibitem{Nose2} S. Nos\'e, A unified formulation of the constant temperature molecular dynamics methods, \text{J. Chem. Phys.} \textbf{81}, 511 (1984).

\bibitem{Hoover} W. G. Hoover, Canonical dynamics: Equilibrium phase-space distributions, \text{Phys. Rev. A} \textbf{31}, 1695 (1985).

\bibitem{Martyna} G. J. Martyna, M. L. Klein, and M. Tuckerman, Nos\'e-Hoover chains: The canonical ensemble via continuous dynamics, \text{J. Chem. Phys.} \textbf{97}, 2635 (1992).

\bibitem{Jang} S. Jang and G. A. Voth, Simple reversible molecular dynamics algorithms for Nos\'e-Hoover chain dynamics, \text{J. Chem. Phys.}~\textbf{107}, 9514 (1997).

%Dipole dipole interaction

\bibitem{Stuhler} J. Stuhler, A. Griesmaier, T. Koch, M. Fattori, T. Pfau,
S. Giovanazzi, P. Pedri, and L. Santos, Observation of dipole-dipole interaction in a degenerate quantum gas, Phys. Rev. Lett. \textbf{95}, 150406 (2005).

\bibitem{Filinov} A. Filinov, and M. Bonitz, Collective and single-particle excitations in two-dimensional dipolar Bose gases, Phys. Rev. A \textbf{86}, 043628 (2012).

%Fermi-Hubbard model

\bibitem{Esslinger} T. Esslinger, Fermi-Hubbard physics with atoms in an optical lattice, Annu. Rev. Condens. Matter Phys. \textbf{1}, 129 (2010).


%free expansion of cold atoms
\bibitem{Dalfovo} F. Dalfovo, S. Giorgini, L. P. Pitaevskii, and S. Stringari, Theory of Bose-Einstein condensation in trapped gases,
Rev. Mod. Phys. \textbf{71}, 463 (1999).

\bibitem{Anderson} M. H. Anderson, J. R. Ensher, M. R. Matthews, C. E. Wieman, and E. A. Cornell, Observation of Bose-Einstein condensation in a dilute atomic vapor, Science \textbf{269}, 198 (1995).

\bibitem{Davis} K. B. Davis et al., Condensation in a gas of sodium atoms, Phys. Rev. Lett. \textbf{75}, 3973 (1995).

%Bose-Hubbard model
\bibitem{Greiner} M. Greiner, O. Mandel, T. Esslinger, T. W. Hänsch, and I. Bloch, Quantum phase transition from a superfluid to a Mott insulator in a gas of ultracold atoms, Nature \textbf{415}, 39 (2002).

%Gaussion interaction
%\bibitem{Mujal} P. Mujal, E. Sarl\'e, A. Polls, and B. Juli\'a-D\'az, Quantum correlations and degeneracy of identical bosons in a two-dimensional harmonic trap, \text{Phys. Rev. A} \textbf{96}, 043614 (2017).

%BKT dipole
%\bibitem{Filinov} A. Filinov, N. V. Prokof’ev, and M. Bonitz, Berezinskii-Kosterlitz-Thouless transition in two-dimensional dipole systems, \text{Phys. Rev. Lett.} \textbf{105}, 070401 (2010).




\end{thebibliography}
\end{document}